\documentclass{article}
\pdfoutput=1

\usepackage{a4wide}
\usepackage{amsmath}
\usepackage{paralist}
\usepackage{booktabs}
\usepackage{makecell}
\usepackage{graphicx}
\usepackage{hyperref}
\usepackage{cleveref}

\title{Impact of etcd deployment on Kubernetes, Istio, and application performance}
\author{Lars Larsson \and William T\"{a}rneberg \and Cristian Klein \and Erik Elmroth \and Maria Kihl}
\date{}
\hypersetup{pdfauthor={Lars Larsson, William T\"{a}rneberg, Cristian Klein, Erik Elmroth, Maria Kihl},pdftitle={Impact of etcd deployment on Kubernetes, Istio, and application performance}}

\begin{document}

\maketitle

\begin{abstract}
By intrinsic necessity, Kubernetes is a complex platform. Its complexity makes conducting performance analysis in that environment fraught with difficulties and emergent behavior. Applications leveraging more ``moving parts'' such as the Istio service mesh makes the platform strictly more complex, not less. In this paper we study how underlying platform constitution and deployment affects application performance, specifically in Kubernetes-based environments. We alter platform constitution via use of native Kubernetes networking or Istio. Platform deployment is altered via etcd data storage location at two extremes on the performance spectrum: network disk and RAM disk. Our results show that etcd performance has a large impact on that of Kubernetes and its ability to perform orchestration actions, and thereby indirectly on the performance of the application. The implication is that systems researchers conducting performance evaluations cannot just consider their specific application as being under test, but must also take the underlying Kubernetes platform into account. To conduct experiments of scientific rigor, we developed an experiment framework for conducting repeatable and reproducible experiments. Our framework and resulting data set are openly available for the research community to build upon and reason about.
\end{abstract}

\section{Introduction}
\label{secIntroduction}

With the worldwide cloud IaaS (infrastructure as a Service) market worth \$32.4 billion\cite{gartner2018market} and a predicted containerization penetration of 75\% by 2022\cite{gartner2019containers}, there is considerable interest within both academia and industry to enable cost-efficient use of the underlying technology. Within container orchestration, Kubernetes has emerged as a de-facto standard\cite{gartner2019containers}, as it simplifies deployment of containerized applications across a cluster of machines in a cloud provider-agnostic way.

Cloud-native applications rely on characteristics such as horizontal scaling and services offered by the underlying platforms, such as service discovery. Interest in offloading application-level networking functionality such as traffic and security token management onto service meshes such as Istio is increasing in the community and becoming a critical element of cloud-native infrastructure\cite{gartner2018servicemesh}.

Containerized platforms and the clouds upon which they are deployed are complex pieces of software, and complexity necessarily impacts performance. Thus, research is needed to identify and quantify sources of performance impact in these systems. Understanding underlying platforms and how their characteristics affect not only their performance and ability to operate correctly, but also the performance of applications deployed onto them is key to achieving cost-effective deployments predictably.

In this paper, \textbf{we study how underlying platform constitution and deployment affects application performance}.
In other words, this paper is concerned with identifying sources of emergent behavior affecting performance that stem from the underlying platform and which are therefore outside of the deployed application's control. To understand the source of this emergent behavior, we provide the technical background in how traffic routing in Kubernetes works, with and without Istio, and the theoretical knowledge needed to perform closed loop load testing. We then present our open source experiment framework, that systems researchers may benefit from in conducting their own performance experiments in these environments. The experiment framework is used to conduct several sets of experiments presented in the paper. One set comprises 1400 experiments, and these indicate the presence of emergent behavior traceable to the underlying Kubernetes and cloud platforms. Thus, we perform experiments that show that etcd is the main culprit, and that its performance can significantly affect that of Kubernetes, and by extension, that of the deployed application.

Our contributions in this paper are as follows:
\begin{itemize}
  \item A thorough explanation of how network traffic is routed among Kubernetes Pods and Services with and without Istio, to our knowledge the first in the research community at this level of detail (Section~\ref{secSystemUnderTest}).
  \item A fully open-source experiment framework, designed for repeatability, reproducibility, recillience, control, data collection, and data exploration (Section~\ref{secExperimentalLabEnvironment}).
  \item A theoretical model and experimental exploration of how to correctly determine cluster sizes of both the system under test and the load generators for performance tests that span across multiple nodes (Section~\ref{secPerformanceAndLoadDimensioning}).
  \item Experimental data showing the impact of etcd performance on the correct functioning of Kubernetes-internal functionality and how this gives rise to emergent behaviors that in turn affect performance of the deployed application (Section~\ref{secEtcdPerformance}).
  \item An open data set comprising several thousands of experiments and millions of measurements used to support the arguments in this paper, ready for further analysis by the research community.
\end{itemize}

Based on our findings, we claim that performance analyses of containerized cloud applications must take the underlying platform into account and consider both application and platform as the system under test.

\section{Networking-focused exploration of Kubernetes}
\label{secSystemUnderTest}

Kubernetes is an orchestrator for containerized applications. As such, it provides functionality required for configuring and deploying applications, automatically managing their life-cycles, service discovery, and managing storage. Of particular interest to us is how network traffic is routed both in native Kubernetes (Section~\ref{secSUTNativeKubernetesDeployment}) and with the service mesh Istio (Section~\ref{secSUTIstioEnabledDeployment}). We are also interested in studying how the amount of self-management orchestration tasks Kubernetes has to do affects application performance (Section~\ref{secSUTResourceLimitations}).

Kubernetes clusters logically consist of sets of master and worker nodes. Master nodes contain the Application Program Interface (API) service, which performs authentication and authorization on a per-operation level. They communicate directly with an etcd database, a distributed key-value store. The master nodes also contain core components such as the scheduler and various Controllers, which automate tasks that for instance, manage the life-cycle of deployed application instances, so called Pods. Pods are ephemeral encapsulations of one or more containers such that they share process namespace, file system, and network context within the particular Pod. Communication between Pods is typically done via the Service abstraction, which provides service discovery and a longer life-cycle than Pods. A group of Pods backing a Service is usually managed by the Deployment Controller and referred to as a Deployment.

The networking abstraction in Kubernetes, Container Network Interface (CNI), provides conceptually simple L2 or L3 networking functionality. For cloud-native applications that require and benefit from higher-level functionality such as routing performed on L7 (application-level) values such as HTTP paths or header values, service meshes act as L7 overlay networks with these additional features. Istio is a service mesh that brings advanced traffic management and observability features and the ability to centrally decide on e.g.\ security policies and rate limitations for the Pods in the mesh. The ability to configure automatic conditional retries for requests (and more) makes Istio popular in the cloud-native community\cite{gartner2018servicemesh}.

How network traffic is routed to members of the Deployment differs depending on whether the native Kubernetes Service abstraction or the Istio service mesh is used. The two approached are detailed in \Cref{secSUTNativeKubernetesDeployment,secSUTIstioEnabledDeployment}, respectively.

In our descriptions of network traffic flows, we assume that the Flannel CNI provider is used, and that it operates in Virtual Extensible LAN (VXLAN) mode. Furthermore, we assume that Kubernetes Services are exposed as NodePort type services.
This is without loss of generality, because even if a cloud provider offers a load balancing service, this is how Kubernetes will expose the service to such load balancers.
Traffic to Services are proxied by kube-proxy, which we assume operated in IP Virtual Server (IPVS) mode. The alternative iptables mode routes traffic via non-deterministic iptables rules instead, but is otherwise conceptually similar.

\subsection{Native Kubernetes deployment \label{secSUTNativeKubernetesDeployment}}

\begin{figure*}[t]
  \includegraphics[width=\textwidth]{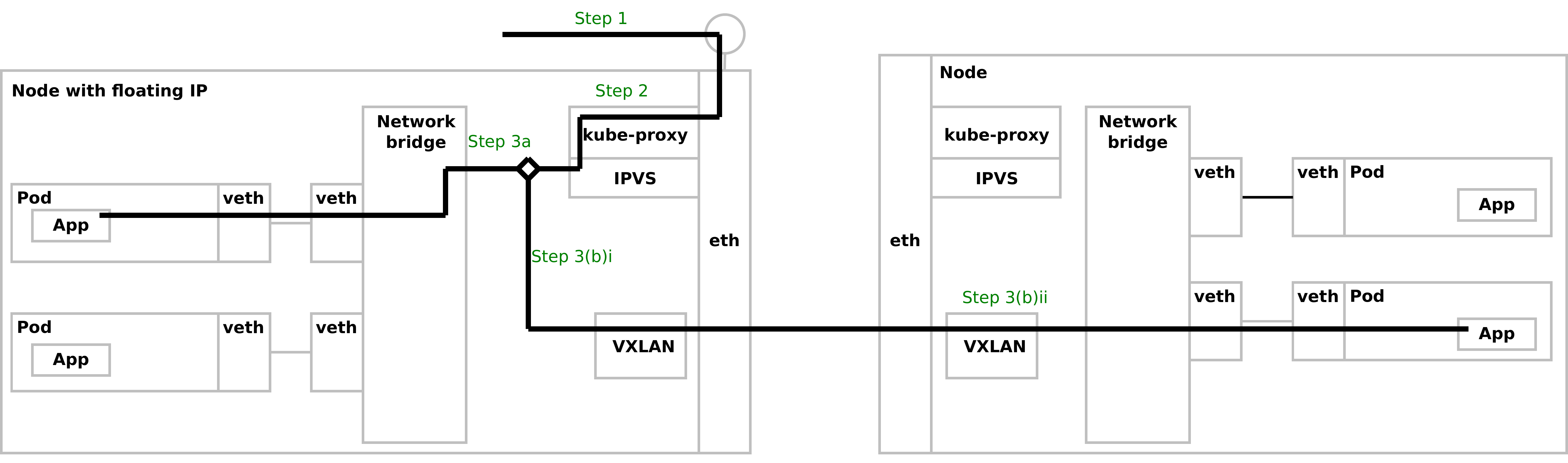}
  \caption{Network traffic flow for an application deployed exposed using a native Kubernetes NodePort Service.}
  \label{figNetworkFlowNativeKubernetes}
\end{figure*}

In native Kubernetes, incoming traffic flows as follows to reach a Pod (cf.\ Figure~\ref{figNetworkFlowNativeKubernetes}):

\begin{enumerate}
  \item \label{networkNativeReachNodePort} The workers's traffic reaches the port specified in the NodePort Service. Any public IP handling via e.g.\ OpenStack floating IP or AWS Elastic IP is performed by Network Address Translation (NAT) by the cloud provider.
  \item \label{networkNativeKubeProxy} kube-proxy listens to the specified port and proxies packets to IPVS in the worker node. IPVS then chooses which Endpoint (i.e.\ Pod) to forward the traffic to, according to its configured scheduler.
  \item Local and remote traffic is treated differently:
    \begin{enumerate}
    \item \label{networkNativeSameNode} If the destination Pod is deployed locally, traffic is routed by IPVS onto a network bridge, where a set of paired virtual Ethernet adapters forward traffic from the host into the Pod.
    \item Alternatively, the destination Pod could be deployed on a different node. Thus, the CNI provider must ensure it reaches the target node.
      \begin{enumerate}
        \item \label{networkNativeFlannelVXLANsend} Flannel's VXLAN mode relies on handling in the Linux kernel, which essentially creates a UDP tunnel between nodes. Other CNI providers operate differently, and may even integrate natively with the virtual networking system used by the cloud provider.
        \item \label{networkFlannelVXLANreceive} The target node unpacks the VXLAN packets and its Flannel forwards them to the target Pod via a network bridge on the target node. Paired virtual Ethernet devices (veth), with one end on the node's network bridge and the other in the Pod, makes sure that the traffic reaches its destination.
      \end{enumerate}
    \end{enumerate}
\end{enumerate}

Matters are kept relatively simple because Kubernetes' network model dictates that no NAT shall be required for node-to-Pod network communication. Thus, a network bridge is used instead. We see, however, that many steps are needed even in this, the arguably simplest case of how to expose a Kubernetes Service (as it does not also involve a separate load balancing service, in addition to the above).

\subsection{Istio-enabled deployment \label{secSUTIstioEnabledDeployment}}

The Istio service mesh is divided in two logical parts: a control plane and a data plane.
Core Istio components manage e.g.\ configuration and security tokens on the control plane level, and configures the underlying data plane with user-specified policies.
Traffic within the data plane flows between Envoy proxies, each deployed as sidecar containers in Pods alongside the main application containers.
These proxies capture all traffic in and out of the Pod, making it controllable from the control plane.

Getting network traffic into the Istio service mesh is done by using an Istio \emph{Ingress Gateway}.
The Ingress Gateway is a Pod that is exposed externally via a Kubernetes Service.
Upon receiving traffic, it uses configuration from the Istio control plane to determine the correct destination within the service mesh.
Such routing can be arbitrarily complex and based on IP addresses and endpoints (as in Kubernetes) or even application-level information, such as HTTP headers and request paths.

A feature of particular interest is that Istio can conditionally issue automatic retries in case of error, e.g.\ only for HTTP status code 500 or above (as they indicate server-side errors). Should such an error be encountered, the Envoy proxy in the Ingress Gateway can be configured to choose a different endpoint up any numer of times, without returning a result to the worker. Error handling of this type is transparent from the worker perspective, and is noticeable only as prolonged waiting time ($t^{\text{resp}}$, cf.\ Section~\ref{secPerformanceAndLoadDimensioning}) for a response.

\begin{figure*}[t]
  \includegraphics[width=\textwidth]{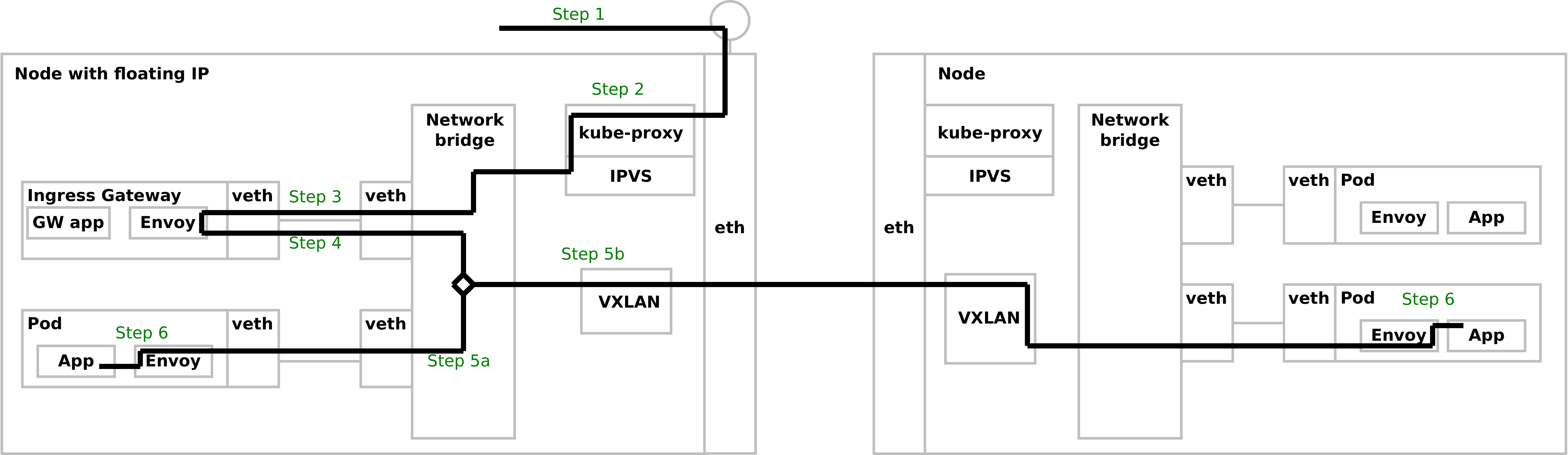}
  \caption{Network traffic flow for an application deployed using the Istio service mesh, where the Ingress Gateway service is exposed as a Kubernetes NodePort service.}
  \label{figNetworkFlowIstio}
\end{figure*}

In contrast to the network traffic flow in native Kubernetes (cf.\ Section~\ref{secSUTNativeKubernetesDeployment}), the one in Istio-enabled deployment includes a number of additional steps.
In the Istio case, we pinned the Ingress Gateway Pod for efficiency to the node with floating IP, and exposed it via Kubernetes NodePort Service. This ensures that at least the initial network hop will always be a local to the same node.
Network traffic flow is as follows (cf.\ Figure~\ref{figNetworkFlowIstio}):

\begin{enumerate}
  \item \label{networkIstioReachGatewayNodePort} As in Native Kubernetes Step~\ref{networkNativeReachNodePort}, but the Service in question is the Istio Ingress Gateway.
  \item \label{networkIstioKubeProxy} As in Native Kubernetes Step~\ref{networkNativeKubeProxy}, but the Service in question is the Istio Ingress Gateway.
  \item \label{networkIstioGateway} As in Native Kubernetes Step~\ref{networkNativeSameNode}, but by design it is known that the Ingress Gateway Pod is on the same host, as it has been pinned to it via a Node Selector.
  \item \label{networkIstioGatewayEnvoy} Envoy uses its configuration (stemming from the Istio control plane) to determine the IP address of an application's service endpoint target Pod. The Envoy proxy sidecar container in the Ingress Gateway Pod has configured iptables to ensure that all Pod traffic is passed through the Envoy proxy.
  \item Local and remote traffic is treated differently:
    \begin{enumerate}
    \item \label{networkIstioSameNode} If the Pod was deployed on the same node, traffic is routed locally to it via the virtual Ethernet adapters and local network bridge. Again, the Envoy proxy sidecar container is the initial recipient of all network traffic.
    \item \label{networkIstioOtherNode} The Pod that shall receive traffic may be deployed on a different node. If so, traffic is passed through a virtual Ethernet tunnel onto a local network bridge and is then passed to the target node via VXLAN, where it is then unpacked, similar to Native Kubernetes Steps~\ref{networkNativeFlannelVXLANsend} and~\ref{networkFlannelVXLANreceive}. A difference is that the recipient is again the Envoy proxy sidecar container in the Pod, as dictated by iptables rules.
  \end{enumerate}
  \item \label{networkIstioEnvoyToPod} Regardless of how traffic got to it, the Envoy proxy sidecar container makes a localhost network call to the main application container.
\end{enumerate}

As the list shows, including Istio to an application's deployment adds \begin{inparaenum}[(a)]
\item an additional network hop and full network stack due to the Ingress Gateway, and
\item additional routing, network stack, and CPU processing due to the use of the Envoy proxy sidecar container.
\end{inparaenum}
The possible benefits of Istio therefore surely cost in terms of additional resources required.
This is a known fact in the Istio community, and their official documentation includes a section on it\footnote{\url{https://istio.io/docs/concepts/performance-and-scalability/}, accessed October 22 2019.}.
How such a performance degradation affects emergent behaviors in the underlying Kubernetes platform, or how performance degradation may be accentuated by them, has to the best of our knowledge not been studied yet in the literature.

\subsection{Interplay of probes and resource limitations\label{secSUTResourceLimitations}}

Kubernetes uses probes to determine if Pods are able to handle incoming requests.
It will at regular (user-defined) intervals probe the containers in each Pod with a configured Readiness Probe. Should one or more container readiness probes fail, the Pod is no longer considered a valid Endpoint of any Service it belongs to, and should therefore no longer receive traffic. This implies that there is a time window where a Pod may be in a state when it is unable to handle traffic, but Kubernetes has not probed it yet. In that case, the HTTP-level response will be 503 Bad Gateway.

Aside from application failure, one reason why a container may fail to respond affirmatively to its readiness probe is that it has been terminated by Kubernetes.
Kubernetes offers the ability to limit the amount of RAM and CPU containers in Pods may use.
Violating the memory limitation by allocating more results in an immediate termination by the Linux Out-Of-Memory (OOM) Killer process.
In contrast, violating CPU is impossible, since it's up to the cgroup functionality in the kernel to limit how much CPU time a process gets scheduled for.
When violations occur, corrective actions have to follow, such as restarting the container.
We can therefore increase the amount of corrective actions that Kubernetes can perform by modifying application deployment so that an application will violate its RAM limit under high load.

While killed and until operational, the container will also fail its readiness and liveness probes, which should affect network traffic routing.
And if routing is not updated in a timely fashion, there should be the aforementioned errors noticeable by workers or load generators.
Note that we do not directly inject faults (cf.\cite{natella2016assessing}). Rather, by increasing the likelihood of containers getting killed, we can study how the underlying Kubernetes (both with and without Istio) platform operates under extreme conditions and how that affects applications deployed onto it.

\section{Open source experiment framework}
\label{secExperimentalLabEnvironment}

In studying both performance and emergent behaviors, \emph{what} to look for is difficult (if not impossible) to know beforehand.
A good data collection and experiment framework therefore needs to yield results that are amenable to post-experiment data exploration.
As one of the key contributions of our work to the systems research community, we present our fully open source experiment framework\footnote{URL withheld until publication}.

Emergent behaviors of Kubernetes clusters manifest themselves in all aspects of performing experiments, not limited to impact on strictly application performance.
They also impacts the ability of the cluster to function at all, e.g.\ due to software, node, or network failures.
Thus, experiments that will be carried out in a Kubernetes environment requires rigorously controlled processes for repeatability, reproducibility (as defined by Feitelson\cite{feitelson2015repeatability}), lest results from one experiment affect another.
Cloud environments in general are time-shared systems with high variance, and thus, cloud performance testing in particular requires repeated experiments and statistical analysis\cite{papadopoulos2019methodological}.
By necessity, yielded data sets are also therefore large.

To this cause, we have developed an experiment framework with the following stated goals:
\begin{enumerate}
	\item \label{featureRepeatabilityReproducibility} experimental repeatability and reproducibility;
	\item \label{featureControl} controlled processes before, during, and after experiments;
	\item \label{featureDataCollection} data collection from various sources; and
	\item \label{featureDataExploration} data exploration facilitation.
\end{enumerate}

This section outlines the design and implementation choices made in this process, and how they relate to the stated goals.

\subsection{Architecture and Components \label{secArchitectureAndComponents}}

From an infrastructure point of view, the main components of our setup are:
\begin{inparaenum}[(a)]
	\item the Kubernetes cluster, upon which the system under test is deployed;
	\item the load generation cluster, which will issue requests toward the system under test and emit test results; and
	\item the experiment framework \emph{Control Node}, which configures, controls, and collects data from experiments.
\end{inparaenum}

Our experiment framework makes no assumption about how Kubernetes is deployed, but does require a way to programmatically reboot nodes as a mitigation if they are marked as ``Not Healthy'' by Kubernetes.

Load is generated using the commonly used Apache JMeter suite.
Apache JMeter is deployed as a distributed system, where each node in the load generation cluster runs JMeter in slave mode.
The load generation cluster is a set of Virtual Machine (VM), possibly running in the same cloud as the Kubernetes cluster.
Crucially, the load generation cluster is \emph{not} running in or on a Kubernetes cluster. It instead uses otherwise minimal VMs, to avoid resource contention affecting the test results.
We have created a systemd Service unit definition for the JMeter slave process, so that its life-cycle can be controlled using standard systemd tools.

The Control Node acts as a one-stop shop to defining, deploying, and tearing down experiments in a controlled way, and it stores all data and metadata related to the experiments. In particular, it has the following components deployed onto it to carry out its operations:
\begin{itemize}
	\item our software to control both the Kubernetes and load generation clusters (Design Goal~\ref{featureControl}) and record metadata about the experiments (Design Goal~\ref{featureRepeatabilityReproducibility});
	\item cloud-specific tooling to allow for rebooting VM, should the need arise (Design Goal~\ref{featureControl});
	\item JMeter master process, which issues commands to the load generation cluster slaves (Design Goal~\ref{featureControl});
	\item data gathering software, which both queries a Kubernetes-deployed Prometheus server and parses JMeter result files (Design Goal~\ref{featureDataCollection});
	\item a PostgreSQL database to store all data (Design Goal~\ref{featureDataExploration}).
\end{itemize}

This separation of concerns and isolation of system under test from the system that performs the tests make this environment suitable for the testing process as described in\cite{edwards2015creating}.

\subsection{Repeatability and Reproducibility \label{secRepeatabilityReproducibility}}

Since performance experiments of cloud software involves many components with emergent behaviors, several repeated experiments with fixed parameters are required.
In our work and experiment framework, we refer to a particular fixation of parameters as a \emph{scenario}.
A particular scenario may, for instance, stipulate that auto-scaling should be used between deployment sizes 10 and 20, and that 120 workers should generate load according to a specific JMeter Test Plan.

For each scenario, researchers define a number of \emph{experiments}. Each experiment has a unique ID, and relates to a particular scenario.
All measurements and metadata will be associated to the experiment they belong to.
We keep track of the status of the experiment, and it will not be marked as \emph{finished} until the Control Node can assert that all postconditions have been fulfilled.
Most crucially among the postconditions is that all data and metadata has been gathered, parsed, and recorded successfully.

If unrecoverable errors occur during an experiment, the experiment will not be marked as finished.
The Control Node will then ensure that the experiment execution will be reattempted at a later time.

The source code of the application under test and the experiment framework themselves are both stored in version controlled Git repositories.
Since code modifications to either may occur during the experimental phase of one's research, part of the metadata stored for each experiment is the Git commit hashes of each repository.
Thus, it is possible to afterward know exactly which experiments were carried out with which version of all relevant software, and it becomes a statistically verifiable judgment call by the researcher if the changes impact the collected data or not.
It also helps with reproducibility (Design Goal~\ref{featureRepeatabilityReproducibility}).

To mitigate the effects of noisy neighbors\cite{nathuji2010qclouds} affecting too many experiments belonging to a single scenario, our experiment framework will choose experiments such that scenarios with few completed experiments will get priority.
If, for instance, the entire suite of scenarios consists of Scenarios A--H, and there are 4 finished experiments for scenarios A--D, but only 3 finished experiments for E--H, an experiment from the E--H set will be selected.
Naively running all e.g.\ 10 experiments of Scenario A, followed by all 10 experiments of Scenario B would run the risk of some noisy neighbor problem affecting a large portion of experiments of any one particular scenario.

\subsection{Controlling Experimental Processes \label{secControllingExperimentalProcesses}}

Controlling the experimental processes is key to ensuring that tests are repeatable and results are reproducible by other researchers.
To that end, our experiment framework performs the following tasks:

\begin{itemize}
	\item refuses to continue unless Git reports no uncommited changes to the working directories of both the system under test and the experiment framework itself;
	\item ensures that a RAM disk is used to store intermittent JMeter results, to avoid additional I/O delays as much as possible;
	\item asserts that all Kubernetes nodes are reporting in as ``Healthy'', restarting them via the cloud's API if this is not the case;
	\item ensures clean process states and lack of Java Virtual Machine (JVM) garbage by re-starting all JMeter slave processes in the load generation cluster;
	\item asserts that Prometheus is accessible, because it will hold performance measurements from the experiment;
	\item asserts that the PostgreSQL database is working and able to store data; and
	\item removes all traces of the system under test from the Kubernetes cluster, including supporting services such as e.g.\ Istio service mesh, and any data stored related to a possible previous run of this experiment to minimize result contamination between experiments\cite{mytkowicz2009producing}.
\end{itemize}

Once these preconditions and processes have been carried out, the environment is ready to deploy the application under test for the experiment.
Meanwhile, metadata such as the starting time is recorded, the application under test is deployed, and the load generation cluster is controlled by the JMeter master process running in the Control Node.

\subsection{Data Collection \label{secDataCollection}}
Data is collected from two main sources: the load generation cluster reports how the system under test behaves from the perspective of an external observer, and a Prometheus system deployed in the Kubernetes cluster reports on the internally observable behavior.

The load generation cluster uses JMeter, which is deployed in distributed mode, using the Control Node as the master of the JMeter cluster.
As previously mentioned, the master process will store any runtime measurement results on a RAM disk to avoid needless disk I/O slowing down the reporting and subsequent load generation of the cluster of slaves.
This is a safe and reasonable approach because if the Control Node should be rebooted during testing, that's considered an unrecoverable error, and the experiment will thus have to be re-run, as per Section~\ref{secRepeatabilityReproducibility}.

The JMeter master configures its slave nodes to buffer large amounts of data points in an effort to reduce additional network traffic overhead to report results to the master node.
Such reporting is costly in terms of network resources, which is arguably one of the most precious resources one would not want to waste during a load generation test.
Additionally, the delay between requests can be drawn from a distribution, and can therefore for example be made to behave like a Poisson process.
Buffers stored in memory are far better for that purpose.

JMeter operates by using a Test Plan encoded as a file, which specifies all aspects of how a single slave should carry out the test.
Typical contents in such a file is delays between requests, whether to operate in open or closed loop fashion, request rates, and simulated user behavior.
The total number of workers (cf.\ Section~\ref{secPerformanceAndLoadDimensioning}) is divided equally among JMeter slaves.
For performance evaluation purposes, the test plan can can be arbitrarily complex, but at least has to be finite: it completes after a pre-determined number of seconds or issued requests.

The Prometheus database is installed and configured by using the Prometheus Operator Helm Chart.
Prometheus is a monitoring solution and database that is enjoying wide-spread adoption in the Kubernetes community.
It operates in a pull fashion, where a master process pulls data from all its configured data sources.
The Prometheus Operator includes such data sources and configuration for effortlessly exposing crucial metrics regarding node and Kubernetes Pod resource usage.
Istio can also function as a pull source.

When an experiment is over, the experiment framework notes the end time in the experiment's metadata in the PostgreSQL database.
Then, all requested metrics are parsed from both the JMeter cluster and from Prometheus.
Metric data resolution is recorded on a per-second level.
All measurements are parsed using purpose-built parsers and subsequently stored in the PostgreSQL database.
To make experiments easy to compare, we normalize time such that timestamps are stored from second 0.
Actual wall-clock time for a particular point in the experiment can be determined by simply adding the timestamp to the recorded start time.

\subsection{Data Exploration \label{secDataExploration}}

Large sets of data are generated when experiments are repeated and wide parameter spaces are explore, particular when metrics are reported with a per-second resolution.
Iteratively understanding such datasets requires the ability to test hypotheses that were not known at the time of data collection.
Thus, the way to work with this data is akin to how data scientists have adopted a workflow centered around data exploration, often interacted with by leveraging e.g.\ Jupyter notebooks.

Our experiment framework stores all data and metadata related to scenarios and experiments in a PostgreSQL database.
This facilitates integration with e.g.\ Jupyter notebooks as well as data analysis libraries such as NumPy and pandas.
By storing data in a relational database rather than in e.g.\ a large set of CSV files, we enable rapid exploration and ability to ask new queries without having to create single-purpose scripts that parse CSV files \emph{en masse}.

The database schema is normalized such that measurements belong to experiments which belong to scenarios.
This makes it trivial to e.g.\ obtain the maximum number of requests served per second for a particular scenario by making the appropriate SQL JOIN statements.

We believe that while e.g.\ pandas DataFrames are able to perform relational algebra operations, the expressive power of SQL far surpasses that offered by pandas.
It is also more efficient, since data does not need to be needlessly read, parsed, transmitted, and represented in main memory if it is irrelevant to the actual query being asked.
However, users that prefer having the full datasets in main memory may request it by making large SQL JOINs to collect it.

\section{Performance and load dimensioning}
\label{secPerformanceAndLoadDimensioning}

  In order to reason about the system under test, a model representation of it is needed.
  Therefore, in this section we parameterize and model the system as described in \Cref{secArchitectureAndComponents}.
  Additionally, a simple set of queueing theory tools to reasoning about how to dimension and load such a system is presented.

  On a high level, the system under test consists of a Kubernetes cluster (here denoted as $\text{k8s}$) with a set of nodes, $\mathcal{N}^{\text{k8s}}$, and a JMeter load generator cluster with a set of worker nodes, $\mathcal{N}^{\text{load}}$.
  An application is hosted on the Kubernetes nodes and is composed of a set of Pods, $\mathcal{P}$.
  These sets are defined in \Cref{eq:kube_nodes,eq:kube_pods,eq:jmeter_workers,eq:jmeter_threads}.

  \begin{align}
  			\mathcal{N}^{\text{k8s}} &= \{n^{\text{k8s}}_{i} \mid i=1,2,...,I\}, \label{eq:kube_nodes}\\
  			\mathcal{P} &= \{p_{j} \mid j=1,2,...,J\}, \label{eq:kube_pods}\\
        \mathcal{N}^{\text{load}} &= \{n^{\text{load}}_{k} \mid k=1,2,...,K\}, \label{eq:jmeter_workers}\\
  			\mathcal{W} &= \{w_{l} \mid l=1,2,...,L\} \label{eq:jmeter_threads}
  \end{align}

  The set of load generating workers in $\mathcal{W}$ are identical and continuously HTTP-request-generating entities that operate in a closed loop.
  In a closed loop system, a $w_{l}$ sends a request, and waits for a response of duration $t^{\text{resp}}=t^{\text{exec}}+t^{\text{rtt}}+Z$ or a time-out $t^{\text{time-out}}$ before proceeding with the next request.
  Note that $t^{\text{time-out}}$ is in the load generator.
  Here $t^{\text{exec}}$ and $t^{\text{rtt}}$ are the expected pod execution time and the round-trip-time between the load generator nodes $\mathcal{N}^{\text{load}}$ and the cluster nodes $\mathcal{N}^{\text{k8s}}$, respectively.
  $Z$ is uncharacterised noise.
  Therefore, a worker $w_{l}$ generates HTTP requests at a rate of $\frac{1}{\min(t^{\text{resp}},t^{\text{time-out}}) + t^{\text{delay}}}$, where $t^{\text{delay}}$ is a user-configurable parameter.
  Decreasing $t^{\text{delay}}$ increases the request rate.
  The total request rate $\lambda$ subjected on $\mathcal{P}$ is thus:

  \begin{equation}\label{eq:request_rate}
    \lambda = \sum^{\mathcal{W}} \frac{1}{\min(t^{\text{resp}},t^{\text{time-out}}) + t^{\text{delay}}}
  \end{equation}

  A pod $p_{j}$ has an expected service rate (HTTP responses/sec) $\mu_{j} = \frac{1}{t^{\text{exec}}+t^{\text{rtt}}+Z}$.
  The service rate for the whole system is therefore, $\mu = \sum^{J} \mu_{j}$.

  The utilisation of $\mathcal{P}$ is defined as $\rho=\lambda/\mu$.
  Note however, since this is a closed loop system, the value of $\rho$ is only valid when $<1$ \cite{kleinrock1975queueing}.
  As such the reasoning in this section will enable us to determine how to overload the system but not to what extent.

  The set of Pods $\mathcal{P}$ in the system are Resource Sharing (RS) and each have a finite queue of an unknown size.
  Here, the set of resources shared are memory, CPU, and I/O.
  A pod $p_{j}$ represents a job or a process.
  It is assumed that resources are dynamically managed and scheduled in each node $n^{\text{k8s}}_{i}$.
  Additionally, even when given the opportunity, applications in such systems do not necessarily consume all resources.
  Therefore, the service rate $\mu_{j}$ for each pod $p_{j}$ is not necessarily linear with resource availability.
  Consequently, the system is represented by a closed $\langle T/M/M/c \rangle-\text{RS}$-type queueing-system, where $c=J$ and $T=L$.
  The Processor Sharing (PS) $\langle T/M/M/c \rangle-\text{PS}$ model presented in \cite{Braband1995} can be seen as basis on which to begin to estimate the properties of such a system.
  However, the actual service rate $\mu^*$ and actual request rate $\lambda^*$ will be less than their theoretical counterparts, $\mu^*\le \mu, \lambda^*\le \lambda$.

  \subsection{Dimensioning the system - an example}
\begin{figure}[t]
      \includegraphics[width=\textwidth]{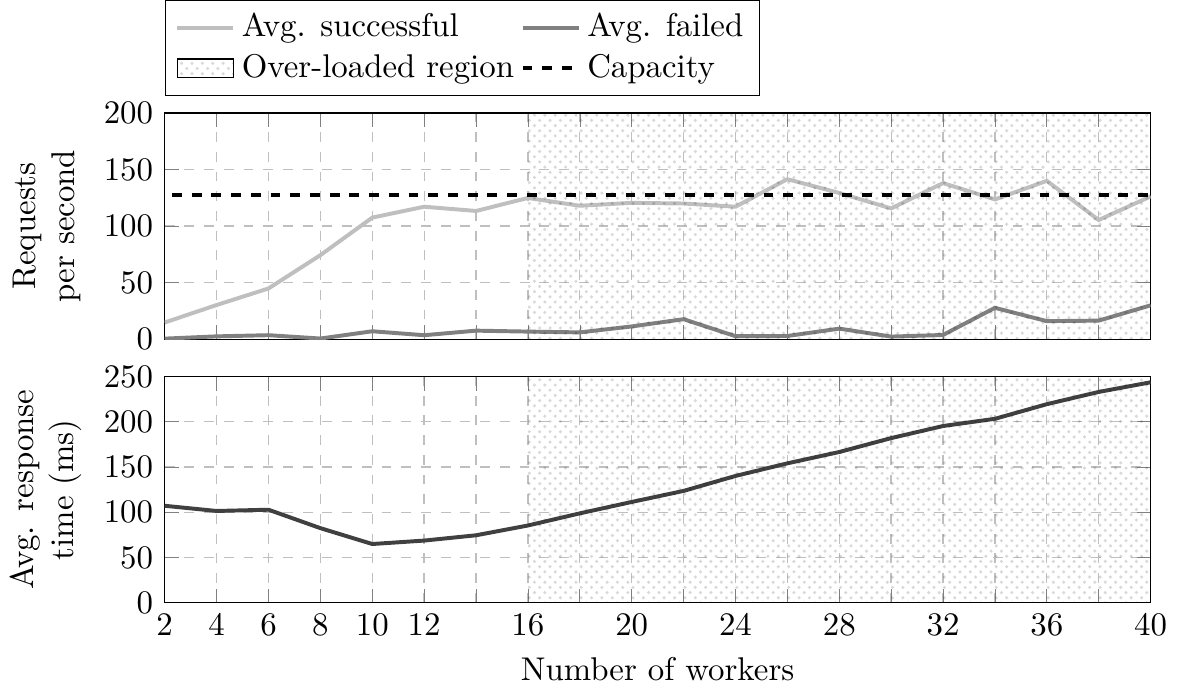}
      \caption{The performance of 20 Pods observed over a range in the number of workers [2,40] and where \protect$t^{\text{time-out}} = \infty$.}
      \label{figLoadGeneration}
    \end{figure}

    There are two fundamental aspects that need to be determined when dimensioning an experimental setup; the capacity of your system $\mu^*$ and to what degree the system is subject to the intended load $\lambda^*$.
    We use the data presented in \Cref{figLoadGeneration} as basis for reasoning about these parameters.
    The scenario\footnote{Scenarios prefixed \texttt{loadgeneration-reduced-workers} in the data set.} depicted in \Cref{figLoadGeneration} is with $I=6, J=20$ and $t^{\text{delay}}=20ms$.
    Furthermore, $L$ is varied in $[2,40]$, in steps of two, with seven repetitions for each value of $L$.
    Note that the objective is to load the system to $\rho \geq 1$.
    The duration of each experiment is 20 min, with seven repetitions.

    The basic performance assumptions are as follows.
    Queueing theory tells us that the successful response rate should plateau at the $L$ where $\rho=1$.
    Around that point, the requests will begin to be either queued or rejected.
    As requests are queued the mean waiting time will be greater than $\min(t^{\text{resp}},t^{\text{time-out}})$.
    This is evident from the lower half of \Cref{figLoadGeneration}, where the waiting time strickly increases with $L$ when $L > 10$.
    Thus, reflected in the upper half of the figure, the system of 20 Pods is not able to produce more than an average of $127.8$ responses per second.

    The best case mean response rate can be estimated from the response time chart in \Cref{figLoadGeneration}.
    The lowest achieved response time ($t^{\text{resp}}$) is $63 ms$.
    Analogously, at most, each worker $w_{l}$ can generate $\approx 12$ requests per second.
    To achieve full utilisation of the $20$ Pods ($I=20$), $\rho=1$, at least $16$ workers are needed ($L=16$), assuming $Z=0$.
    As this is a closed system, $t^{\text{resp}}$ is representative of the service time for a pod when $\rho \le 1$.
    The expected service rate per pod is then $\approx 16$ responses per second.
    Note that the dip can be attributed to node level scheduling prioritisation decisions.
    Again, the system under test does not strictly adhere to theory.
    However, $\rho=1$ is somewhere in $L=[10,16]$ but guaranteed at $L>16$.
    This discrepancy is due to noise ($Z$) from the policy events, execution scheduling, and difficult-to-predict circumstances.
    Nevertheless, using the queuing model above allow us to reason about the properties of the system, and estimate its performance bounds.
    Additionally, in most scenarios we want to operate the system in the over-loaded region $\rho \geq 1$, which we can now derive.

\section{Experimental setup}
\label{secExperiments}

In this paper, we study the effects that the constitution and deployment of the underlying Kubernetes (and Istio) platform has on application performance.

  The application under test must be \emph{simple} enough that interesting behavior can be correctly attributed to the surrounding environment into which it is deployed, yet sufficiently \emph{complicated} that it consumes a realistic set of computational resources.

  \subsection{Kubernetes cluster and cloud infrastructure \label{secKubernetesClusterCloudInfrastructure}}
    Motivated by reasons of both the possibility to gain additional operational insight that one will not get from a public cloud provider and that of lower cost, we used a private OpenStack-based cloud for our experiments.
    This particular private cloud was built for research purposes with a heavy focus placed on fault-tolerance.
    A design decision in that cloud is therefore that all disk access is actually backed by a Ceph cluster of redundant network-attached storage nodes.
    As a consequence, there is no truly local instance storage.
    This property can have a significant impact and, as Section~\ref{secExperience} shows, is the cause of some emergent behavior.
    However, as Section~\ref{secExperience} also shows, the emergent behavior lets us gain valuable insight related to Kubernetes and application performance.

    Kubespray was used to install an, at the time when experimentation started, stable version of Kubernetes (1.12.6). The cluster consisted of:
    \begin{itemize}
      \item 1 master node with a floating IP attached, 8GiB of RAM, 4 VCPUs. The master node is a dedicated master (does not permit application Pods to be scheduled), and also includes the etcd database.
      \item 1 worker node with a floating IP attached, 4GiB of RAM, and 2 VCPUs.
      \item 5 worker nodes without floating IPs attached, with the same amount of resources as the other worker.
    \end{itemize}

    With respect to networking, we used the Flannel CNI provider, and the Kubernetes cluster was configured to use the, according to official documentation, higher performance IPVS implementation of Services, rather than the iptables-powered one. Our setup closely follows the one shown in Figures~\ref{figNetworkFlowNativeKubernetes} and~\ref{figNetworkFlowIstio}, in that it does not use an cloud-provided load balancing service. This is state of the art for OpenStack, with only 2 of 22 public providers offering support for the Octavia load balancing service\footnote{\url{https://www.openstack.org/marketplace/public-clouds/}, accessed October 15, 2019.}. We used version 1.1.5 of the Istio service mesh.

  \subsection{Application under test \label{secApplicationUnderTest}}
    The application under test in our experiments is an intentionally simple single-tier stateless function that performs an image manipulation task.
    A request to the application specifies a specified rectangle.
    The application, encodes and returns the rectangle as a Base64-encoded PNG image.
    The application is \emph{simple} in the sense that a single-tier application does not have multiple layers of inter-component communication over a network, and is therefore not delayed by e.g.\ blocked upstream network connections.
    Furthermore, emergent behavior for such a simple application should largely be due to the Kubernetes and cloud infrastructure.
    Yet it is sufficiently \emph{complicated} because it consumes a realistic amount of I/O (loading a specified image), CPU resources (selecting the rectangle and encoding the result), and memory (holding the image and any intermediary results), in addition to, of course, servicing requests over the network.

    In common practice web application fashion, we used the Flask web framework for Python 3 to serve over HTTP.
    Furthemore, image manipulation is done using pillow and Base64 encoding by the Python 3 standard library.
    The application is packaged as a Docker image and, to make testing simple, includes a predefined set of images.
    The application has been made fully open source and is available online\footnote{\url{URL withheld until publication}}.

    \subsubsection{Application deployment \label{secSUTDeployment}}
      The application is deployed as a Kubernetes Deployment.
      To make it possible to provoke errors and corrective actions by Kubernetes, the Pod specification in the Deployment includes resource limitations.
      These, and the intended effects such limitations have during load testing, are described in more detail in Section~\ref{secSUTResourceLimitations}.
      Resource limitations are also required for the subset of our scenarios that make use of the Horizontal Pod Autoscaler (HPA) to scale the size of the Deployment based on CPU load.

      Recalling Section~\ref{secSUTResourceLimitations}, we aim to increase the likelihood of corrective actions being required by Kubernetes.
      Thus, in our experiments, we limit the amount of RAM allocated to our main application container.
      Under low load, the limit should have no effect, but under high load, the Python garbage collector is not fast enough to de-allocate memory used to handle previous requests, and the process eventually uses more memory than it is allowed.
      Determining this limit is obviously application-dependent, and in our case, 64MiB of RAM was experimentally determined to provide the sought effects.
      Rather than deterministically terminating containers e.g.\ after X requests or after X seconds of CPU load over a given threshold, we chose this approach because we do not want to control any emergent behaviors, and thereby risk causing them to disappear. Instead, we want to study them as they occur.

Our application, when deployed on top of Istio, defines an Istio \emph{Virtual Service} which includes an application-level routing rule.
Additionally, our application defines that Istio should issue automatic retries in case of error, should the error be of HTTP status 500 or above (cf. Section~\ref{secSUTIstioEnabledDeployment}).

\section{Experiments on Istio vs.\ native Kubernetes traffic routing} 
\label{secShowCase}

To verify and demonstrate our experimental infrastructure and experiment framework, we conducted a basic experiment comprising of 1400 experiments over 320 scenarios\footnote{Scenarios prefixed \texttt{scaling} in the data set.}, where we vary the number of Pod instances $J$, workers $L$, and network deployment (native Kubernetes or Istio service mesh).
The experiments were designed to study the effects, if any, of using Istio instead of native Kubernetes traffic routing. As well as to observe the standard deviation in the resulting data.
In particular when combined with auto-scaling.
The reader should see this section as a demonstration for what can be done using our experiment framework.

As shown in Sections~\ref{secSUTNativeKubernetesDeployment} and~\ref{secSUTIstioEnabledDeployment}, use of Istio adds both additional work in terms of network stacks to traverse and additional routing to be carried out.
But how much, and do we get any benefits from it that could improve performance overall anyway?

The load generation cluster is configured to generate a consistent load level in a closed-loop manner, as defined in Section~\ref{secPerformanceAndLoadDimensioning}.
Scenarios define that we vary the number of initial Pods to be used for auto-scaling ($J\in[1,2,3,4,5,10,15,20]$).
To reduce the experiments running time and because we expect the successful request rate to plateau with large rnumber of initial Pods, the experiment's resolution is reduced at over 5.
The upper limit for the HPA is fixed to $\hat{J}=20$ Pods for all scenarios.
Our scenarios also define a varying number of workers that generate load, ranging from $K \in [1,\ldots,20]$ per JMeter slave.
With 6 JMeter slaves, the total number of workers $L$ is thus $6,12,\ldots,120$.
Additionally, the workers are configured with $t^{\text{delay}}=20ms$, where $t^{\text{delay}}$ is the mean of a poisson process, and $t^{\text{time-out}}=\infty$.
Reflective of \Cref{figLoadGeneration}, the experiments spans over both under loaded $\mu < 1$ and over loaded regions $\mu \geq 1$.
Finally, the duration of each experiment is 20 min.

\subsection{An experiment on Istio's performance}
\begin{figure}[t]
  \centering
  \includegraphics[width=\textwidth]{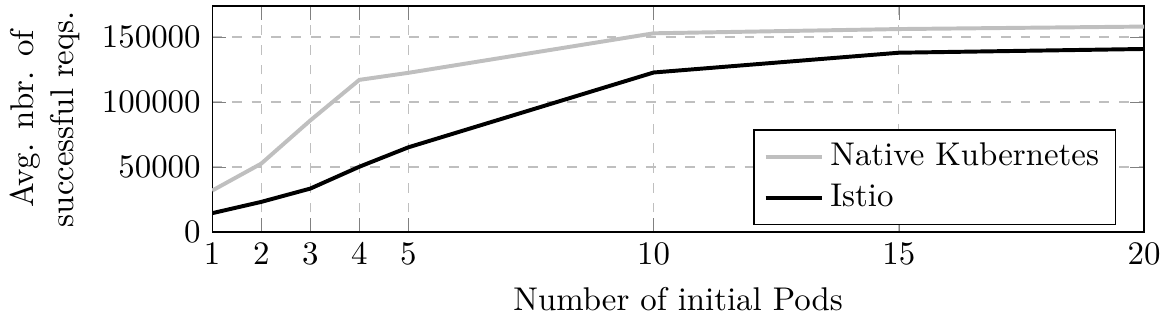}
  \caption{Average successfully handled requests as a function of number of initial Pods in an auto-scaled suite of experiments.}
  \label{figSuccessfulRequests}
\end{figure}

Figure~\ref{figSuccessfulRequests} shows the average successfully served requests as a function of initial Pods.
As hypothesized, the performance of Istio is lower than that of native Kubernetes, in terms of throughput, since it carries out more work per request (cf.\ Section~\ref{secSUTIstioEnabledDeployment}).
A native Kubernetes deployment is strictly more performant than an Istio deployment.
The discrepancy is at most more than two-fold.
To be more precise, a statistical analysis of the entire dataset shows that there is a \emph{significant difference} between using Istio and native Kubernetes networking ($p=8.8\times10^{-33}$, as per Kruskal-Wallis H-test for independent samples).


\subsection{Noisy and missing data}
\begin{figure}[t]
  \centering
  \includegraphics[width=\textwidth]{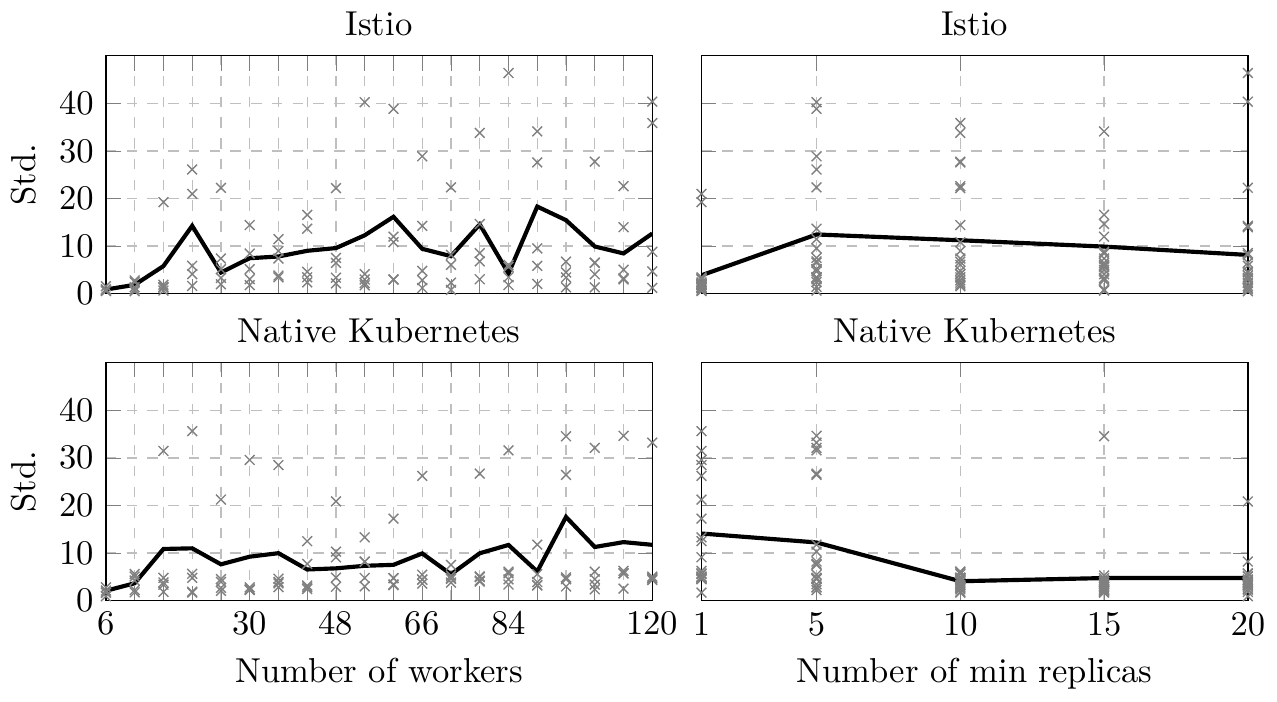}
  \caption{Standard deviation in successful requests per second for each scenario. Xs respresent datapoints for each scenario, while the solid line represents the mean over scenarios. Values further from zero indicate lack of stability and thus presence of emergent behavior.}
  \label{figDifferenceSuccessfulResponses}
\end{figure}

The results from these initial experiments also showed high variability within each scenario.
Variability is a reflection of noise $Z$ in the system.
Figure~\ref{figDifferenceSuccessfulResponses} shows the standard deviation in successfully handled requests per second for each scenario.
Small values indicate stability, lack of emergent behavior, and that reality behaves as theory claims.
Alas, the data instead shows that as the number of ``moving parts'' in our system increases (Pods $\mathcal{P}$, workers $\mathcal{W}$, use of Istio), so does the variability in our results.
It is also evident that in a under loaded system $L<16$ the standard deviation is significantly lower.

Another data quality problem is that Prometheus, with its pull model, is not guaranteed to be able to pull data from its configured sources. Should a data source be unable to reliably buffer its data until it gets scraped, gaps in the data may occur. Many applications bundle a Prometheus instance, since Kubernetes does not provide a central one for the entire cluster. This may become problematic when it comes to locating the data after an experiment. The experiment framework presented in Section~\ref{secExperimentalLabEnvironment} therefore supports multiple Prometheus instances.

\section{Emergent behaviors in Kubernetes due to etcd performance impact}
\label{secExperience}

\begin{figure}[t]
  \centering
  \includegraphics[width=\textwidth]{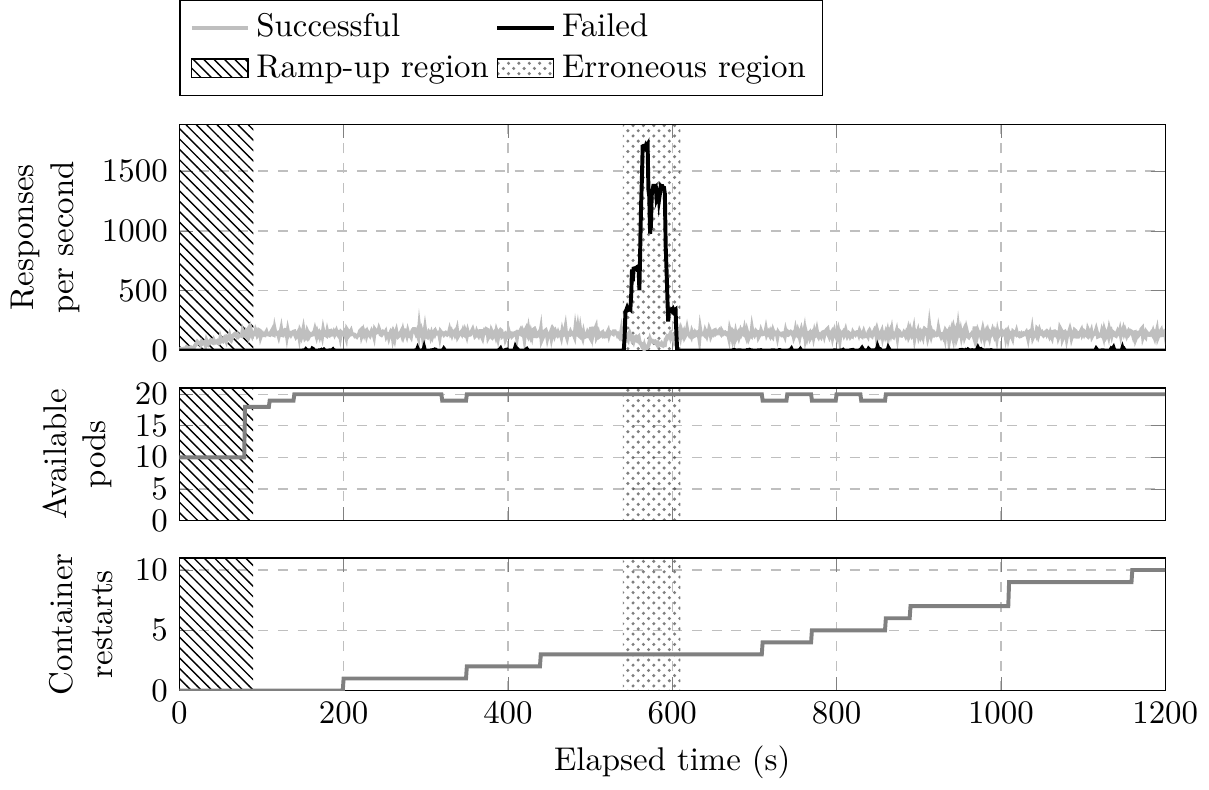}
  \caption{An experiment where Kubernetes failed to take corrective action during an extended period of time (experiment ID 2c0609c0-9ef7-11e9-9783-4b7ce873c223).}
  \label{figErroneousRegion}
\end{figure}

\begin{figure}[t]
  \centering
  \includegraphics[width=\textwidth]{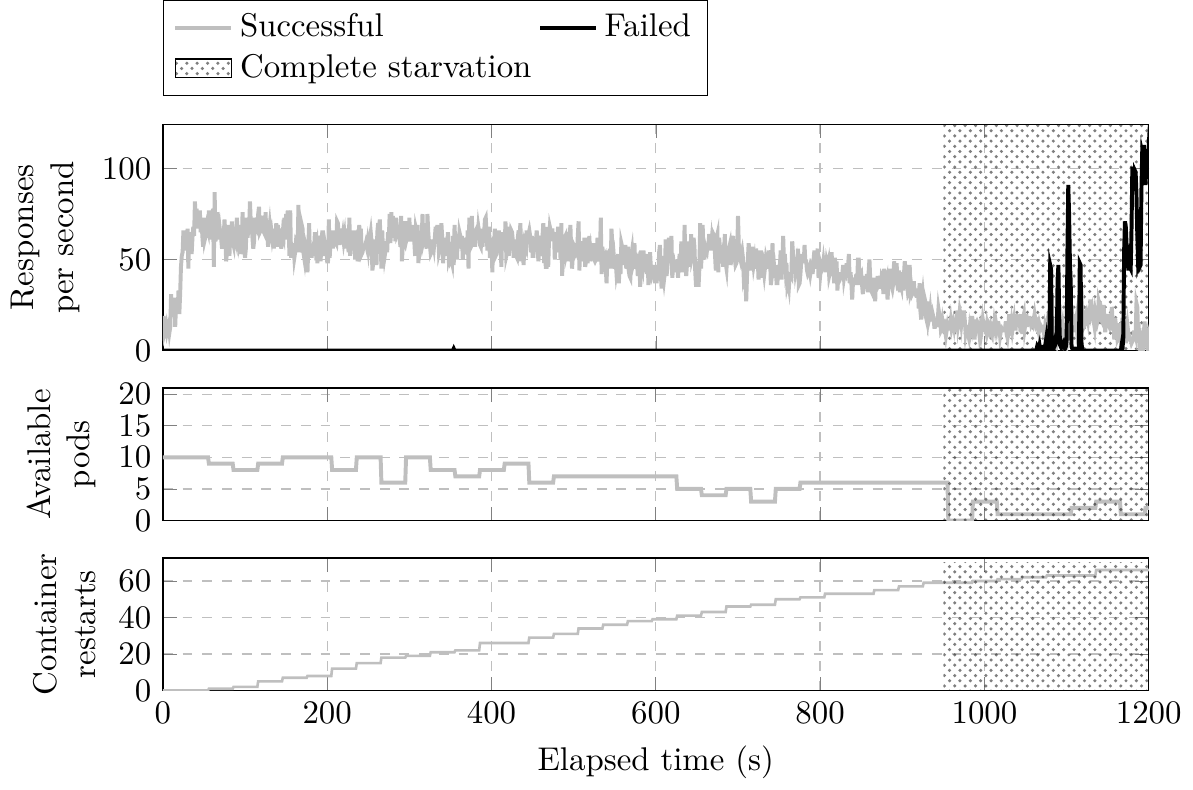}
  \caption{Compounded failures leading to starvation and the HPA unable to alleviate the situation (experiment ID bbebcc1a-9ef6-11e9-82fc-870f50601c23).}
  \label{figHpaFailure}
\end{figure}

The large variability in results shown in Section~\ref{secShowCase} and Figure~\ref{figDifferenceSuccessfulResponses} inspire deeper study of time series data.
Two archetypes of errors are of particular interest, showcased by example experiments in Figures~\ref{figErroneousRegion} and~\ref{figHpaFailure}. These experiments inherit the same basic parameters as those in \Cref{secShowCase}. In the first (Figure~\ref{figErroneousRegion}), it is apparent that \emph{some} problem occurs and that the system is unable to perform corrective action for a while.
In the marked erroneous region in \Cref{figErroneousRegion}, the  successful response rate plummets and the failure rate increases rapidly, but is limited by $\lambda^{\text{max}}=6000$ (the maximum of \Cref{eq:request_rate} given $t^{\text{delay}}=20ms$).
Consequently, application performance suffers and failures occur due to incorrect routing (cf.\ Section~\ref{secSUTResourceLimitations}).
The second error archetype shows that frequent errors renders the HPA unable to request the correct number of desired Pods (Figure~\ref{figHpaFailure}).

Realizing that these emergent behaviors cast a shadow of doubt over our results from these initial experiments, we devoted our attention to determine the root cause. Investigation along various paths ultimately lead to etcd Linux systemd journal entries, which showed a pervasive presence of warnings about operations taking longer than expected.

Recall that, at its core, Kubernetes consists of a database (etcd), an API server, and a set of components and controllers that carry out orchestration tasks.
Such tasks and corrective actions are specified and triggered by data made accessible via the API, and stored in the etcd database.
Our research presented in this section shows that this dependence on etcd has considerable performance impact.
With all other parameters being as equal as a cloud environment can provide, etcd performance affects that of the application, in particular under high load.
We are not aware of any other scientific work that details this relationship, and believe that performance tests that do not take this into account run a high risk of producing wrong data without doing anything obviously wrong\cite{mytkowicz2009producing}.

\subsection{Factors impacting etcd performance \label{secEtcdPerformance}}

The etcd database is a distributed key-value store that utilizes the Raft algorithm for distributed consensus. To ensure that results are stable and persistent, it makes extensive use of disk I/O.
It is known in the etcd community that poor I/O per second performance will negatively affect etcd\footnote{\url{https://github.com/etcd-io/etcd/blob/master/Documentation/platforms/aws.md}, accessed October 22 2019}.
How such poor performance affects Kubernetes and the applications deployed onto it, however, has not been published to the best of our knowledge.
With regard to etcd deployment, operators must choose one of two alternatives, neither of which are entirely satisfactory:

\begin{enumerate}
\item either etcd can use only fast instance storage that is local to the VM, which shares the same life-cycle as the VM, which thus makes it susceptible to data loss or service disruption if the VM is terminated; or
\item etcd uses a network-attached block storage disk, which offers greater availability and independence from the life-cycle of any VM it is attached to, but significantly poorer performance.
\end{enumerate}

Other than the backing storage medium, of course CPU and network capacity affect performance as well.
The dominant factor, unless networks are truly saturated or greatly under-provisioned, is however the disk I/O performance.
We focus only on the dominant disk I/O factor, and it turns out to be trivial to modify with significant impact.

\subsection{Experiments on etcd deployment performance impact on that of the application \label{secEtcdApplicationPerformanceInterplay}}

To determine the possible impact of etcd performance on that of our application, we devised a suite of scenarios (cf.\ Section~\ref{secRepeatabilityReproducibility}).
In one set of scenarios, we let etcd use a network-attached block storage disk (recalling Section~\ref{secKubernetesClusterCloudInfrastructure}, our private cloud provider did not provide local instance storage). In another set, we let etcd use a RAM disk as its data storage. This is \emph{highly} inadvisable outside of research, as the loss of etcd data will render etcd and thus Kubernetes unable to recover. But its I/O performance is second to none!

\begin{figure*}[t]
  \includegraphics[width=\textwidth]{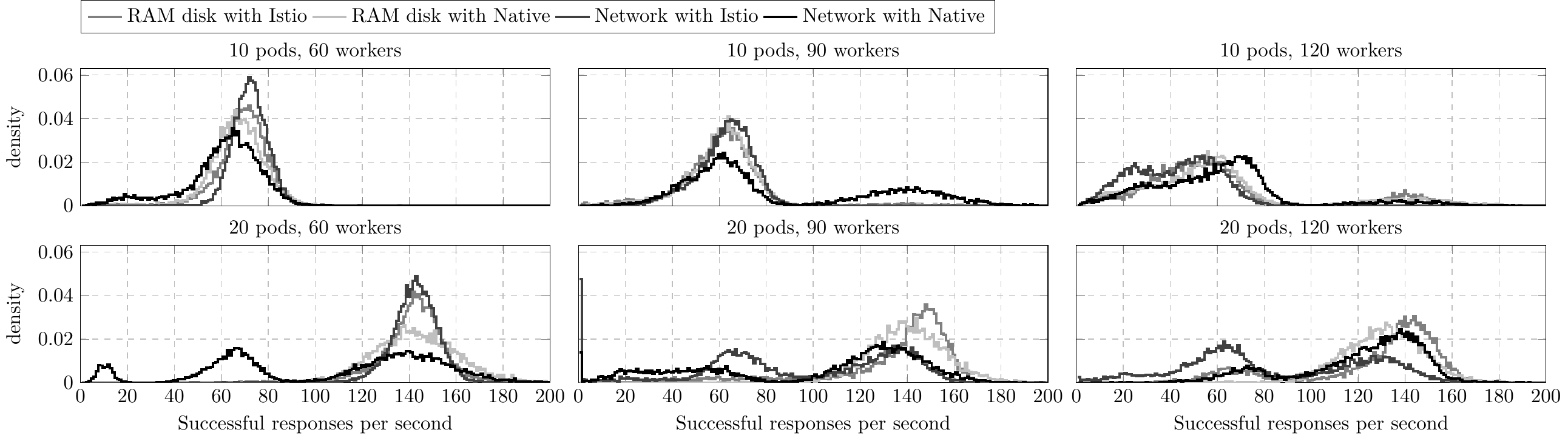}
  \caption{Application performance impact from etcd deployment, presented as normalized density plots of successfully serve worker requests.}
  \label{figEtcdPerformanceImpact}
\end{figure*}

The scenarios define experiments with a fixed number of Pods ($J \in [10,20]$) and with a fixed number of workers per JMeter worker node $L \in [10, 15, 20]$.
The only parameters under our control is how etcd is deployed: with RAM disk or network-attached storage.
Results of these experiments on the application's ability to successfully serve requests are shown in Figure~\ref{figEtcdPerformanceImpact}.
Statistical analysis of the obtained results is given in Table~\ref{tabEtcdPerformanceImpact}.
Because the data does not have a normal distribution, the statistical analysis uses the Kruskal-Wallis H-test for independent samples instead of the T-test over all measurements of successful requests per second for all experiments belonging to the different scenarios.

\begin{table}[t]
  \caption{Statistical analysis of application performance impact as a function of etcd deployment. \textbf{All} comparisons show that there is a \emph{statistically significant difference} (\protect$p<0.05$) between using RAM disk or not. Presented p-values are calculated using Kruskal-Wallis H-test for independent samples over all successful requests per second measurements for all experiments belonging to the different scenarios.}
  \label{tabEtcdPerformanceImpact}
  \centering
  \begin{tabular}{cccc}
    \toprule
    \thead{Pods} & \thead{Workers} & \thead{Native Kubernetes \\ RAM disk significant?} & \thead{Istio \\ RAM disk significant?} \\
    \midrule
    10 & 60 & \makecell{($p=4.07\times10^{-115}$) } & \makecell{($p=7.36\times10^{-90}$)} \\
    10 & 90 & \makecell{($p=3.48\times10^{-137}$) } & \makecell{($p=5.37\times10^{-36}$)} \\
    10 & 120 & \makecell{($p=1.68\times10^{-20}$) } & \makecell{($p=2.02\times10^{-284}$)} \\
    20 & 60 & \makecell{($p=0$) } & \makecell{($p=0$)} \\
    20 & 90 & \makecell{($p=0$) } & \makecell{($p=0$)} \\
    20 & 120 & \makecell{($p=2.21\times10^{-37}$) } & \makecell{($p=0$)} \\
    \bottomrule
  \end{tabular}
\end{table}

\subsubsection{Interpretation of results \label{secInterpretationEtcdPerformanceResults}}

From Figure~\ref{figEtcdPerformanceImpact} and Table~\ref{tabEtcdPerformanceImpact}, it can be determined that when etcd performance suffers, so does that of Kubernetes.
And in turn, as that data shows, so does any deployed application.
Key Kubernetes functionality depends on a rapidly responding and functional etcd, including but not limited to storage and retrieval of:

\begin{itemize}
  \item readiness and liveness probe results,
  \item Service Endpoint objects,
  \item scheduling decisions,
  \item measurements that the HPA bases its actions on, and
  \item leader election for components such as the Kubernetes Scheduler and Controller Manager.
\end{itemize}

We have identified three serious problems directly related to poor etcd performance.
When the system is pushed to its limits, some or all of these problems can manifest, resulting in failures such as shown in \Cref{figHpaFailure,figEtcdPerformanceImpact}.
These problems are:

\begin{enumerate}
  \item Slow etcd performance makes the set of Service Endpoints slow to update, causing kube-proxy to route traffic to the container that failed its probes. The response code is then HTTP 503 Bad Gateway.
  \item Control loops in Kubernetes in e.g.\ its Scheduler and Controller Manager fail to acquire leadership leases, which force them to exit and restart. They are unable to resume operation until they have acquired a leadership lease, and therefore, no new decisions from them will be made to help resolve ongoing problems.
  \item A container that cannot report performance measurements for an interval of time will fail to cause the HPA to react.
\end{enumerate}

The third problem state may manifest as a consequence of the first two, and the effect is that other Service Endpoints in a Deployment get higher than expected load, and the HPA does nothing to resolve the situation.
If more containers in the Service Endpoint Pods start to fail, the effect is a downward spiral of failures.
Figure~\ref{figHpaFailure} shows what this looks like, with the Deployment unable to scale up from the 10 initial Pods, and instead, the entire Deployment gets overwhelmed and gets no additional Pods from the HPA.
Ultimately, the Deployment is down to just a fraction of what it should have been (other experiments in the same scenario scaled up to 20 Pods).

\section{Conclusions} 
\label{secConclusions}

Our work shows that while Kubernetes ships with powerful abstractions that make deployment easier across cloud providers, it adds a substantial amount of ``moving parts''. These moving parts cause emergent behavior, in particular when under stress from high load. This impacts performance evaluations. Our data indicates that at the core of the problem is the etcd database and its performance. We have not found results in the literature that describes the relationship between poor etcd performance and the impact it can have on the application and the system as a whole. The implication of our findings is that performance tests on Kubernetes-based cloud environments need to take into account how etcd is deployed, and whether it has sufficient resources and sufficiently fast I/O to perform its work. Thus researchers are encouraged to consider not only their application but also the Kubernetes platform as their system under test. 

We presented our experiment framework, which aids researchers in running Kubernetes-based performance tests in a controlled manner that is repeatable and enables data exploration via tools common in the data science community. We believe this to be a great workflow for exploration and hypothesis verification. It is our hope to encourage and enable the community to collect more large data sets upon which research can be built\cite{schroeder2010failures,tan2012prepare,yabandeh2009crystalball}.

We have also shown, in detail, how network traffic is routed through both native Kubernetes Services and the Istio service mesh. The choice of network data plane also affects the number of moving parts that gives rise to emergent behavior. Our data shows, however, that while Istio does cost in terms of resources, it is able to give a better overall performance for highly loaded systems in terms of successfully served requests.

\bibliographystyle{plain}
\bibliography{refs}

\begin{thebibliography}{10}

\bibitem{Braband1995}
Jens Braband.
\newblock Waiting time distributions for closed m/m/n processor sharing queues.
\newblock {\em Queueing Systems}, 19(3):331--344, Sep 1995.

\bibitem{edwards2015creating}
Sarah Edwards, Xuan Liu, and Niky Riga.
\newblock Creating repeatable computer science and networking experiments on
  shared, public testbeds.
\newblock {\em SIGOPS Oper. Syst. Rev.}, 49(1):90--99, January 2015.

\bibitem{feitelson2015repeatability}
Dror~G. Feitelson.
\newblock From repeatability to reproducibility and corroboration.
\newblock {\em SIGOPS Oper. Syst. Rev.}, 49(1):3--11, January 2015.

\bibitem{gartner2018servicemesh}
{Gartner Inc.}
\newblock Innovation insight for service mesh.
\newblock Market report, 2018.

\bibitem{gartner2018market}
{Gartner Inc.}
\newblock Market share analysis: {IaaS} and {IUS}, worldwide, 2018.
\newblock Market report, 2018.

\bibitem{gartner2019containers}
{Gartner Inc.}
\newblock Best practices for running containers in production.
\newblock Market report, 2019.

\bibitem{kleinrock1975queueing}
Leonard Kleinrock.
\newblock {\em Queueing Systems: Volume 1: Theory}.
\newblock Wiley-Interscience, 1975.

\bibitem{mytkowicz2009producing}
Todd Mytkowicz, Amer Diwan, Matthias Hauswirth, and Peter~F. Sweeney.
\newblock Producing wrong data without doing anything obviously wrong!
\newblock In {\em Proceedings of the 14th International Conference on
  Architectural Support for Programming Languages and Operating Systems},
  ASPLOS XIV, pages 265--276, New York, NY, USA, 2009. ACM, ACM.

\bibitem{natella2016assessing}
Roberto Natella, Domenico Cotroneo, and Henrique~S Madeira.
\newblock Assessing dependability with software fault injection: A survey.
\newblock {\em ACM Computing Surveys (CSUR)}, 48(3), 2016.

\bibitem{nathuji2010qclouds}
Ripal Nathuji, Aman Kansal, and Alireza Ghaffarkhah.
\newblock Q-clouds: Managing performance interference effects for qos-aware
  clouds.
\newblock In {\em Proceedings of the 5th European Conference on Computer
  Systems}, EuroSys '10, pages 237--250, New York, NY, USA, 2010. ACM, ACM.

\bibitem{papadopoulos2019methodological}
Alessandro~Vittorio Papadopoulos, Laurens Versluis, Andr{\'e} Bauer, Nikolas
  Herbst, J{\'o}akim Von~Kistowski, Ahmed Ali-eldin, Cristina Abad,
  Jos{\'e}~Nelson Amaral, Petr T\r{u}ma, and Alexandru Iosup.
\newblock Methodological principles for reproducible performance evaluation in
  cloud computing.
\newblock {\em IEEE Transactions on Software Engineering}, 2019.

\bibitem{schroeder2010failures}
Bianca Schroeder and Garth~A. Gibson.
\newblock A large-scale study of failures in high-performance computing
  systems.
\newblock {\em IEEE Transactions on Dependable and Secure Computing},
  7(4):337--350, Oct 2010.

\bibitem{tan2012prepare}
Yongmin Tan, Hiep Nguyen, Zhiming Shen, Xiaohui Gu, Chitra Venkatramani, and
  Deepak Rajan.
\newblock Prepare: Predictive performance anomaly prevention for virtualized
  cloud systems.
\newblock In {\em 2012 IEEE 32nd International Conference on Distributed
  Computing Systems}, pages 285--294. IEEE, 2012.

\bibitem{yabandeh2009crystalball}
Maysam Yabandeh, Nikola Knezevic, Dejan Kostic, and Viktor Kuncak.
\newblock Crystalball: Predicting and preventing inconsistencies in deployed
  distributed systems.
\newblock In {\em The 6th USENIX Symposium on Networked Systems Design and
  Implementation (NSDI’09)}. USENIX, 2009.

\end{thebibliography}

\end{document}